\documentclass[prl,twocolumn,showpacs,superscriptaddress,amsmath]{revtex4}
\usepackage{graphicx}
\usepackage{amssymb}
\usepackage{citesort}

\begin{document}

\title{Organic superconductors: the need to go beyond effective\ $\frac{1}{2}$-filled band models}
\author{N. Gomes}\affiliation{Department of Physics, University of
  Arizona, Tucson, AZ 85721} 
\author{R. T. Clay }\affiliation{Department of Physics and Astronomy and HPC Center for
  Computational Sciences, Mississippi State University, Mississippi
  State, MS 39762} 
\author{S. Mazumdar}\affiliation{Department of
  Physics, University of Arizona, Tucson, AZ 85721} \date{\today}
\begin{abstract}
A frustrated, effective $\frac{1}{2}$-filled band Hubbard-Heisenberg
model has been proposed to describe the strongly dimerized
charge-transfer solid families $\kappa$-(ET)$_2$X and
Z[Pd(dmit)$_2$]$_2$. In addition to unconventional superconductivity
these materials also exhibit antiferromagnetism, candidate spin-liquid
phases, and in the case of Z=EtMe$_3$P, a so-called valence-bond solid
phase.  We show that neither superconductivity nor the valence-bond
solid phase occurs within the Hubbard-Heisenberg model, indicating
that the effective $\frac{1}{2}$-filled band model is unsuitable for
these materials.
\end{abstract}

\maketitle 

Low-dimensional organic charge transfer solids (CTS) are being
intensively studied because of their many unusual competing and
coexisting electronic phases.  The most studied among them are
probably the $\kappa$-(ET)$_{2}$X and Z[Pd(dmit)$_2$]$_2$ families,
which, depending on the anion X$^-$ or cation Z$^+$ exhibit
unconventional superconductivity (SC), N\'eel antiferromagnetic (AFM)
order, charge ordering (CO), candidate quantum spin liquid (QSL)
behavior, and valence-bond solid (VBS) order \cite{Kanoda11a}. The
apparent similarity between these with the cuprate superconductors
have been noted by many investigators. The semiconductor-SC transition
in the CTS occurs under the application of pressure at constant
carrier density, which suggests that the transition is driven by a
small modification of a particular parameter of an appropriate
Hamiltonian. The key questions then are, what is the minimal model,
and which is the parameter whose changes give the competing phases.

Experimental observations appear to give a simple answer to these
questions.  The 2:1 (1:2) stoichiometry of $\kappa$-(ET)$_{2}$X
(Z[Pd(dmit)$_2$]$_2$), indicates 0.5 hole (electron) per cationic
(anionic) molecule. The crystal structures consist of dimers of
molecules arranged in an anisotropic triangular lattice.  The strong
role of repulsive electron-electron (e-e) Coulomb interactions is
indicated by the observation of N\'eel AFM in
$\kappa$-(ET)$_2$Cu[N(CN)$_2$]Cl, a candidate QSL phase in
$\kappa$-(ET)$_2$Cu$_2$(CN)$_3$, and a VBS phase in
EtMe$_3$P[Pd(dmit)$_2$]$_2$ \cite{Kanoda11a}. Taken together, the {\it
  effective} $\frac{1}{2}$-filled band Hubbard model, or some variant
of it, now appears to be the appropriate minimal model, with the dimer
units as the sites. The anisotropy of the triangular lattice, {\it
  i.e.}, the degree of frustration, is the variable parameter that
changes under external pressure or internal pressure effect caused by
counter-ions with large size. Such a picture readily explains the
observed commensurate AFM at large anisotropy. Motivated by Anderson's
resonating valence bond (RVB) theory \cite{Anderson73a} many
investigators have proposed that the QSL and VBS phases can be
explained within the correlated effective $\frac{1}{2}$-filled band
scenario. For moderate Hubbard $U$, an AFM semiconductor-paramagnetic
metal (PM) transition occurs with increasing frustration.  D-wave SC
mediated by fluctuations of the AFM ordering at the AFM-PM boundary
has been also proposed based on mean-field and dynamic mean-field
theories (DMFT)
\cite{Schmalian98a,Kino98a,Kondo98a,Vojta99a,Kyung06a,Gan05a,Powell,Rau11a}.

\begin{figure}[tb]
\includegraphics[width=3.2in]{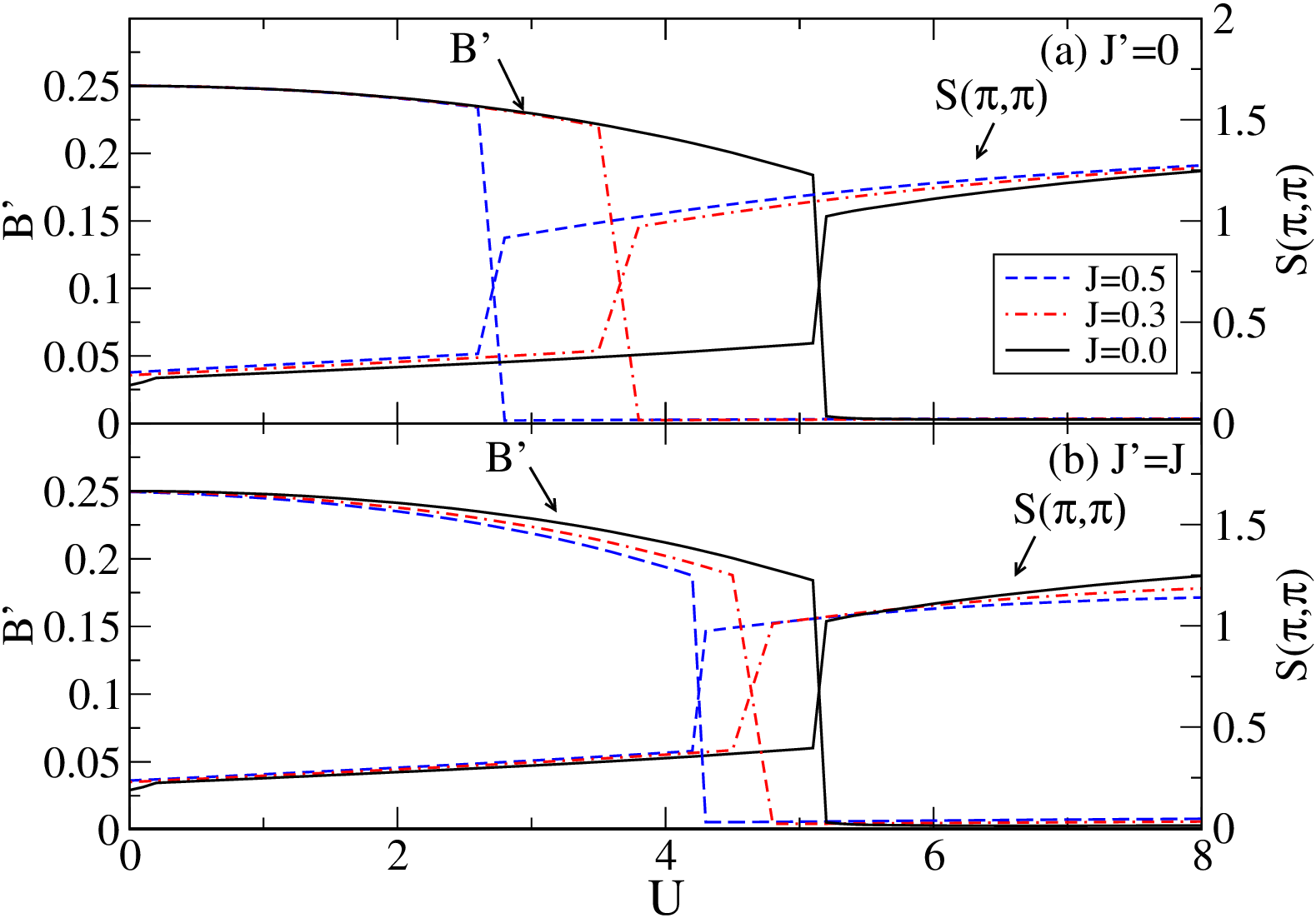}
\caption{(color online) Diagonal bond order $B'$ and spin structure
  factor S($\pi,\pi$) versus $U$ for $t'$=0.4. (a) $J'=0$ (b)
  $J'=J$. Points are calculated at a spacing of $\Delta U$=0.1 The $J$
  interaction strengthens the AFM phase.}
\label{bofig1}
\end{figure} 

Numerical calculations have, however, failed to find SC within the
triangular lattice $\frac{1}{2}$-filled band Hubbard model
\cite{Clay08a, Dayal12a, Tocchio09a}. Numerical studies have also
failed to find a VBS phase in the model \cite{Watanabe03b}.  Although
the $\frac{1}{2}$-filled Hubbard model on the anisotropic triangular
lattice does not appear to support SC, closely related models continue
to be suggested as the appropriate theoretical model for describing
the SC transition in the CTS.  It has been claimed that the simple
Hubbard model does not include all the spin-spin interactions that
play an important role in the CTS, and that additional spin exchange
unrelated to the Hubbard $U$ must be incorporated to correctly capture
AFM fluctuation effects \cite{Gan05a,Powell,Rau11a}.  This has
led to theoretical works on the so-called Hubbard-Heisenberg model
given below.  The goal of this Letter is to critically examine whether
the addition of a Heisenberg exchange term, assumed to be independent
of the Hubbard onsite interaction $U$, causes a superconducting phase
to occur in the $\frac{1}{2}$-filled band anisotropic triangular
lattice model.  A second objective is to see whether the combined
effects of $U$, $J$, and frustration can mediate the VBS phase found
in the EtMe$_3$P[Pd(dmit)$_2$]$_2$ \cite{Tamura09a}.

We consider the following Hamiltonian,
\begin{eqnarray}
 H&=&-t\sum_{<ij>}B_{i,j}-t'\sum_{\{ij\}}B_{i,j}
+U\sum_{i}n_{i\uparrow}n_{i\downarrow}\nonumber\\
&+& J\sum_{<ij>}\vec{S_{i}}\cdot\vec{S_{j}} + 
J'\sum_{\{ij\}} \vec{S_{i}}\cdot\vec{S_{j}}.\label{ham}
\end{eqnarray}
In Eq.~\ref{ham} sites $\langle ij\rangle$ are nearest neighbors on a
square lattice while sites $\{ij\}$ are the next-nearest neighbors
across a diagonal of each square plaquette.
$B_{i,j}=\sum_\sigma(c^\dagger_{i,\sigma}c_{j,\sigma}+H.c)$, where
$c^\dagger_{i,\sigma}$ 
creates an electron of spin $\sigma$ on site $i$;
$n_{i\sigma}=c^\dagger_{i\sigma}c_{i\sigma}$.  $\vec{S}_i$ is the
(spin-$\frac{1}{2}$) spin operator for site $i$.  All energies will be
given in units of $t$.  We limit our analysis here to the region of
lattice anisotropy appropriate for the title materials, $t'\alt 1$
\cite{Kandpal09a,Nakamura09a,Scriven12a}.

We consider two limiting cases: (i) $J'=0$, when the added Heisenberg
interactions do not frustrate N\'{e}el antiferromagnetism, and (ii)
$J'=J$, which frustrates the AFM state.  The $J'=0$ limit was studied
by Gan et al using renormalized mean-field theory \cite{Gan05a} and
the $J'=J$ limit by Rau and Kee using a slave-rotor mean-field theory
\cite{Rau11a}.  Powell and McKenzie studied variable $J'$ $[0<J'/J<2
]$ using an RVB ansatz \cite{Powell}.  SC was found in some region of
the parameter space by all of these authors.  Several authors also
found a spin-liquid phase \cite{Powell,Rau11a}.  Rau and Kee examined
the isotropic lattice and in addition found a VBS phase \cite{Rau11a} that was
claimed to explain the VBS in EtMe$_3$P[Pd(dmit)$_2$]$_2$
\cite{Tamura09a}.

Here we examine the ground state of Eq.~\ref{ham} using exact
diagonalization of a 4$\times$4 lattice.  Exact diagonalization was
previously used to study the model in the limit $J=0$, concluding that
no SC or enhancement of the pairing correlations by $U$ is present
\cite{Clay08a}. Despite the small lattice size (even smaller lattices
are however used in cluster DMFT calculations \cite{Kyung06a}) the
validity of these 4$\times$4 results has been confirmed by recent Path
Integral Renormalization Group (PIRG) \cite{Kashima01b} calculations on considerably
larger lattices\cite{Dayal12a} that arrived at even stronger
conclusions regarding the absence of SC.

\begin{figure}[tb]
\includegraphics[width=3.2in]{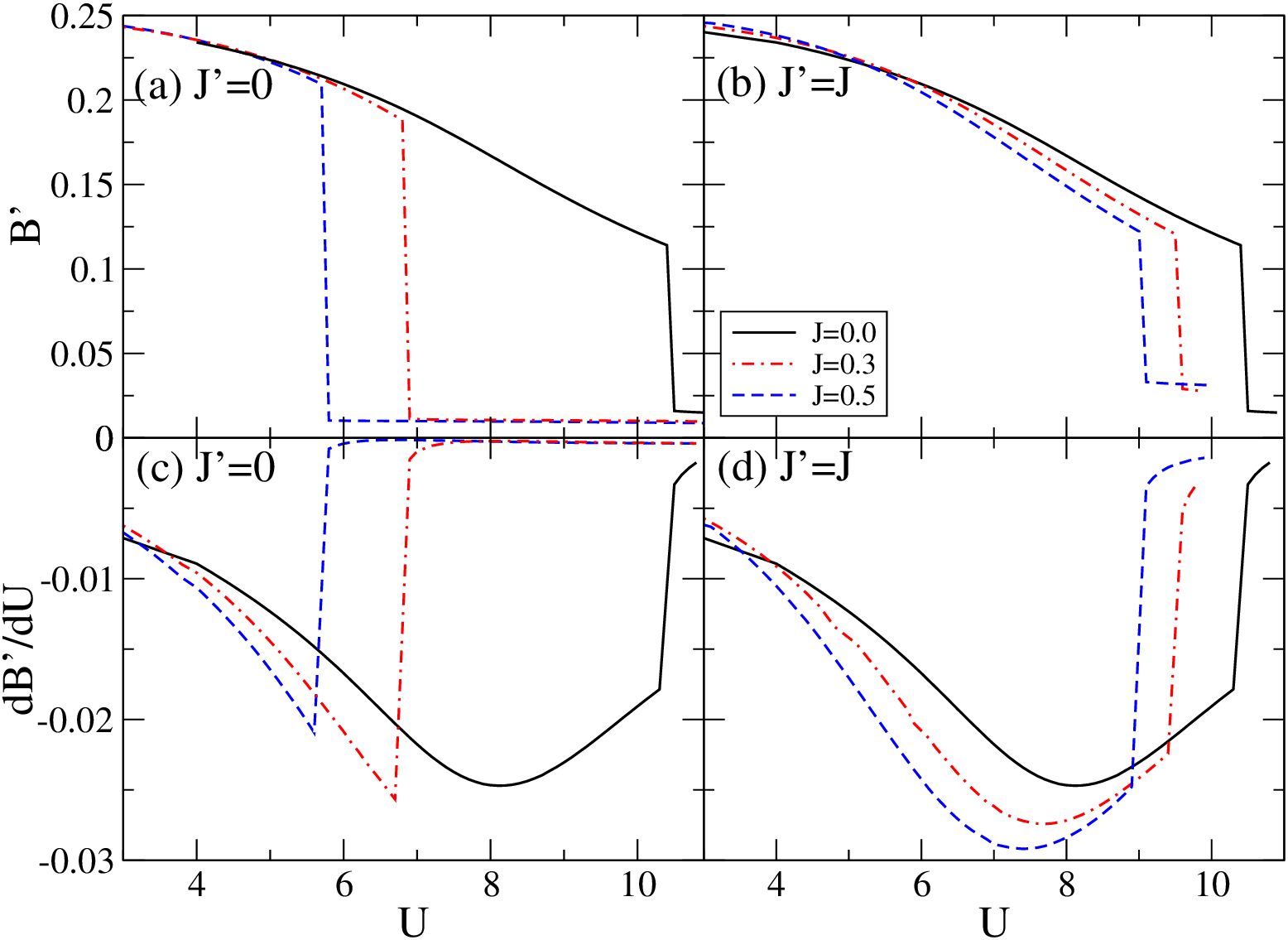}
\caption{(color online) Diagonal bond order $B'$ versus $U$ for
  $t'=0.8$ with (a) $J'=0$, and (b) $J'=J$. The derivative of $B'$,
  $dB'/dU$ with (c) $J'=0$ and (d) $J'=J$.  $B'$ is calculated for
  points separated by $\Delta U=0.1$. For $J'=0$ no NMI phase is
  found. When $J'\neq 0$, both PM--NMI and NMI--AFM phase boundaries
  shift to smaller $U$. The solid, dashed-dotted and dashed curves
  correspond to $J$=0.0, 0.3, and 0.5, respectively.}
\label{bofig2}
\end{figure}
It will be useful to briefly recall the ground-state phase diagram in
the $J=J'=0$ limit
\cite{Kashima01a,Morita02a,Watanabe03b,Mizusaki06a,Koretsune07a,Clay08a,Dayal12a,Tocchio09a,Yoshioka09a}.
Known ground state phases include two AFM phases with N\'eel and
120$^o$ order, a PM phase and a gapless non-magnetic insulator (NMI)
or QSL phase
\cite{Kashima01a,Morita02a,Watanabe03b,Mizusaki06a,Koretsune07a,Clay08a,Dayal12a,Yoshioka09a}.
The NMI phase is found between the AFM and PM phases for $t'\agt 0.5$
\cite{Kashima01a,Morita02a}.  Near the isotropic lattice ($t'\approx
t$) magnetically ordered states with ${\bf Q}\neq(\pi,\pi)$ are found
for large $U$ \cite{Mizusaki06a,Clay08a,Yoshioka09a}. However, such
non-N\'{e}el AFM ordering is not found experimentally
\cite{Kanoda11a}.
 
We first consider the $t'\alt0.5$ region of the phase diagram where a
direct PM--AFM transition is found for $J=J'=0$.  We have calculated
the diagonal bond order  $B'\equiv\langle B_{i,j} \rangle$ for
 sites $i$ and $j$ connected by the $t'$ bond
and the spin structure factor
\begin{equation}
S({\bf Q})=\frac{1}{N}\sum_{j,k}e^{i{\bf Q}\cdot({\bf r}_j-{\bf
  r}_k)} \langle
(n_{j,\uparrow}-n_{j,\downarrow})(n_{k,\uparrow}-n_{k,\downarrow})\rangle.
\end{equation}
Fig.~\ref{bofig1}(a) and (b) show $B'$ 
and $S(\pi,\pi)$ versus $U$ for $t'=0.4$ and several
different $J$ (0, 0.3, 0.5). A sudden drop in $B'$ at $U=U_{c}$,
simultaneous with a sudden increase in $S(\pi,\pi)$, indicates
transition to an insulating AFM state \cite{Morita02a,Clay08a}.  For
both $J'=0$ and $J'=J$, $J$ lowers the $U_c$ for transition to AFM,
broadening the AFM region. As shown in Fig.~\ref{bofig1} the largest
broadening occurs when $J'=0$.

As $t'$ increases the size of the discontinuity in $B'$ and
$S(\pi,\pi)$ decreases.  This is due to the appearance of a NMI phase
in between the PM and AFM phases for $t'\agt 0.5$
\cite{Kashima01a,Morita02a}.  While the transition between PM and AFM
phases for $t'\alt 0.5$ is discontinuous as $U$ is varied, for larger
$t'$ the PM--NMI transition at $U_{c1}$ and the NMI--AFM transition at
$U_{c2}$ are continuous \cite{Morita02a}.  The transitions are now
determined by examining the derivatives of order parameters with
respect to $U$ \cite{Clay08a}. In Fig.~\ref{bofig2} we plot $B'$
versus $U$ and its derivative $dB/dU$ for $t'=0.8$, calculated using a
centered-difference approximation with a $U$ grid of $\Delta U=0.1$.
We find that for $J'=0$ (Fig.~\ref{bofig2}(a) and (c)), the stability
of the AFM phase is enhanced by $U$; $U_{c1}$ decreases with $J$.  At
the same time, the NMI phase is suppressed: the inflection point in
$B'$ in Fig.~\ref{bofig2}(a) for $J=0$ and the corresponding minimum
in $dB'/dU$ in Fig.~\ref{bofig2}(c) disappear even for quite small
values of $J$.  For $J'=J$ (Fig.~\ref{bofig2}(b) and (d)), the system
progresses through two transitions, PM--NMI and NMI--AFM, as in the
pure Hubbard model.  As seen in Fig.~\ref{bofig2}(b) and (d), there is
a slight narrowing of the NMI phase.
\begin{figure}[tb]
\includegraphics[width=3.2in]{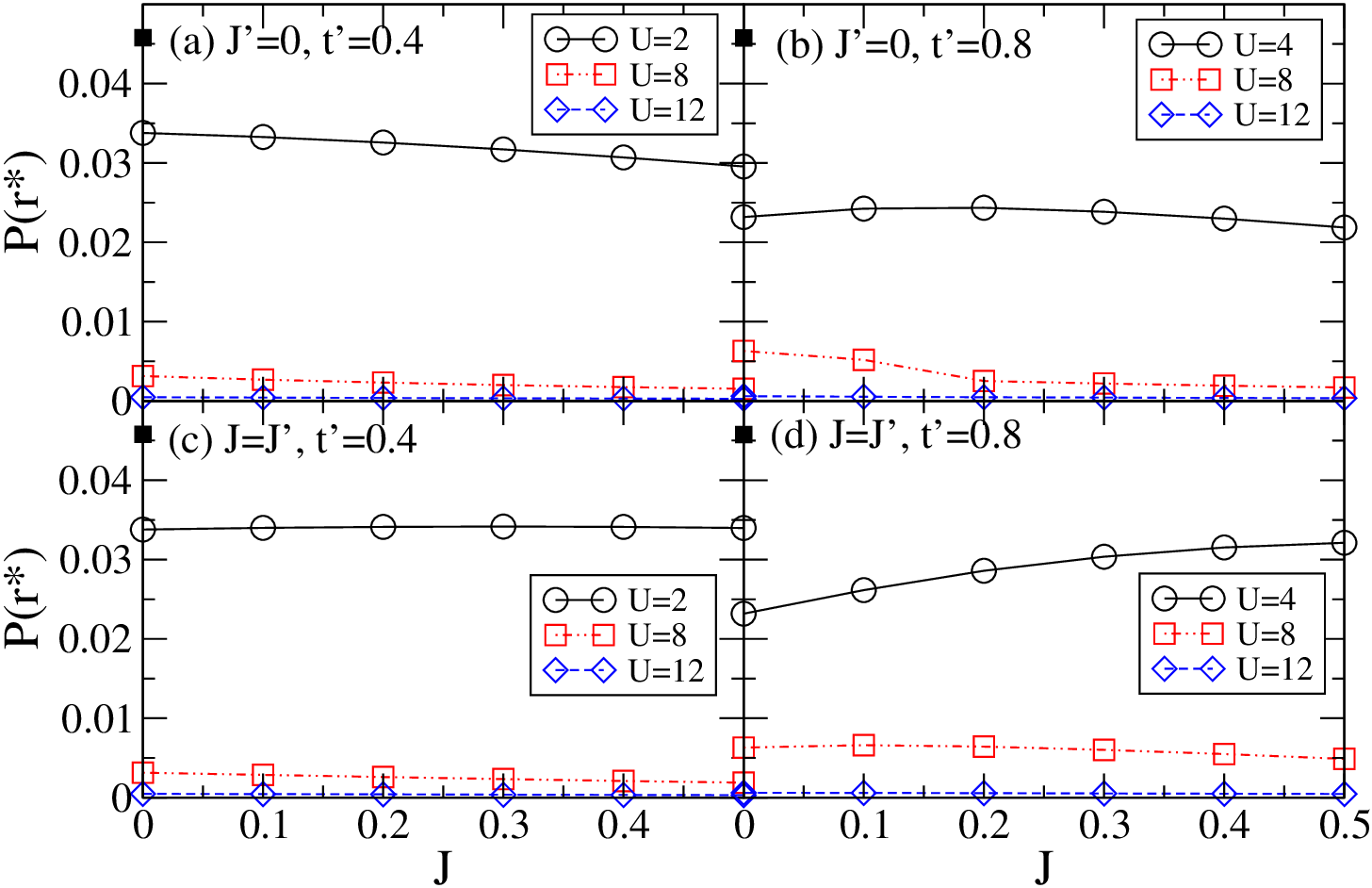}
\caption{(color online) Long-range $d_{x^2-y^2}$ pair-pair correlation
  function $P(r^\star)$ (see text) as a function of $J$ for (a)
  $t'$=0.4, $J'$=0 (b) $t'$=0.8, $J'$=0 (c) $t'$=0.4, $J'$=$J$, and
  (d) $t'$=0.8, $J'$=$J$. $P(r^\star)$ at $U=J=0$ is
  shown by the filled square on each plot.  Independent of the value
  of $J$, long-range pair-pair correlations decrease monotonically as
  $U$ increases.}
\label{scfig}
\end{figure}

A necessary condition for SC is that the pair-pair correlation
function $P(r)$ for pairs of appropriate symmetry reaches a constant
value as $r\rightarrow\infty$.  An additional requirement for SC
mediated by interactions is that the pair-pair correlations are
enhanced by the interaction \cite{Clay08a}. In exact diagonalization
studies of the $J=0$ model, $P(r)$ for all symmetries was found to
decrease monotonically with $U$ from the $U=0$ limit
\cite{Clay08a}. Large-lattice studies found further that {\it as a
  function of distance $r$}, the magnitude of $P(r)$ decreased with
distance faster than the $U=0$ solution \cite{Dayal12a}.  These two
results indicated that SC is not present in the $J=0$ limit.  Here we
focus on any possible enhancement of $P(r)$ due to $J$.

We calculate the pair-pair correlation function as a function of
distance, $P(r)=\langle \Delta^\dagger_i\Delta_{i+\vec{r}} \rangle$,
where pair-creation operators  $\Delta^\dagger_i$ are defined as
\begin{equation}
\Delta^\dagger_i= \frac{1}{\sqrt{2}}\sum_{\vec{\nu}} g(\vec{\nu}) 
(c^\dagger_{i,\uparrow}c^\dagger_{i+{\vec{\nu}},\downarrow}
- c^\dagger_{i,\downarrow}c^\dagger_{i+{\vec{\nu}},\uparrow}).
\label{pair}
\end{equation}
The phases $g(\vec{\nu})$ determine the symmetry of the superconducting
pairs. In our calculations we considered $s$, $d_{x^2-y^2}$, and
$d_{xy}$ pairing symmetries \cite{Clay08a}.  Out of these pairing
symmetries, we found that for $t'<t$, $P(r)$ is largest for
$d_{x^2-y^2}$ pairing symmetry.

Although we have calculated $P(r)$ for all $r$ we show here our
results for the largest possible $r$, $r^\star$.  Fig.~\ref{scfig}
shows the pair-pair correlation $P(r^\star)$ as a function of $J$ for
$d_{x^2-y^2}$ pair symmetry.  Representative values of $U$ are chosen
in each panel to correspond to the different regions of the phase
diagram (PM, AFM, NMI).  In Fig.~\ref{scfig}(a) and (b) we take $J'=0$
and two different values of $t'$, 0.4 and 0.8. In both
Figs.~\ref{scfig}(a) and (b) a direct PM--AFM transition exists.  As
with $J=0$ \cite{Clay08a}, $P(r^\star)$ decreases monotonically with
$U$, with its value decreasing discontinuously at the PM--AFM
transition. As Figs.~\ref{scfig}(a) and (b) show, $P(r^\star)$ also
decreases monotonically with $J$ for $J'=0$.  The primary effect of
$J$ here is to expand the AFM region (Fig.~\ref{bofig1}), which {\it
  decreases} the potential phase space available for SC.  In
Fig.~\ref{scfig}(c)-(d) we take $J'=J$. In Fig.~\ref{scfig}(c),
$t'=0.4$ and again there are only PM and AFM phases. The behavior is
very similar to the $J'=0$ case. Fig.~\ref{scfig}(d) is for $J'=J$ and
$t'=0.8$, with $U$ values chosen to represent points in the PM, NMI,
and AFM phases. In the PM region for large $t'$ ($t'\agt 0.7$) there
is a small increase of $P(r^\star)$ with $J$, but $P(r^\star)$ remains
considerably below its uncorrelated value.  Pairing correlations
continue to decrease monotonically with $U$.

Summarizing our results for the $d_{x^2-y^2}$ pair-pair correlations,
for $t'\alt0.7$, $J$ and $J'$ decrease $P(r)$. In the larger $t'$
region, although there is a weak enhancement of $P(r)$ compared to
$J=J'=0$, the pair correlations never exceed the value obtained for
the uncorrelated limit.

The compound EtMe$_3$P[Pd(dmit)$_2$]$_2$ exhibits a spin gap below 25K
\cite{Tamura09a}.  The low temperature phase has been described as a
VBS by the original investigators \cite{Tamura09a}.  Pressure induced
VBS-SC transition, - analogous to the AFM-SC and QSL-SC transitions in
the $\kappa$-(ET)$_2$X - occurs in EtMe$_3$P[Pd(dmit)$_2$]$_2$.  Any
candidate model that is valid for CTS with dimerized units should
therefore have VBS order in some region of the phase diagram.  Further
motivation to find VBS phases in models of interacting electrons comes
from the extensive recent theoretical investigations of
frustration-driven AFM-VBS quantum phase transitions within quantum
spin models \cite{Jiang12a}.  The VBS phase, if it at all appears
within the present model, should appear in the highly frustrated
region of the phase diagram. This is in agreement with the estimation
of $t'/t\approx$0.9 in EtMe$_3$P[Pd(dmit)$_2$]$_2$, obtained from
ab-initio calculation \cite{Scriven12a}. In the context of the present
model the parameter region of interest is large $U$ with $t'\sim t$,
in between the N\'eel and 120$^\circ$ AFM phases.  Previous PIRG
numerical studies \cite{Watanabe03b} as well as more recent work
\cite{Tocchio13a} on the Hubbard ($J'=J=0$) model have, however, found
the NMI rather than VBS in this region.

As with SC, the VBS phase has been claimed within slave-rotor theory
for the isotropic Hubbard-Heisenberg model ($t'=t$, $J'=J$) for both
$J=0$ and $J>0$ \cite{Rau11a}.  The $J$ interaction was found to
strengthen the VBS order \cite{Rau11a}. We therefore investigate the
Hubbard-Heisenberg Hamiltonian in this highly frustrated region.
\begin{figure}[tb]
\includegraphics[width=3.2in]{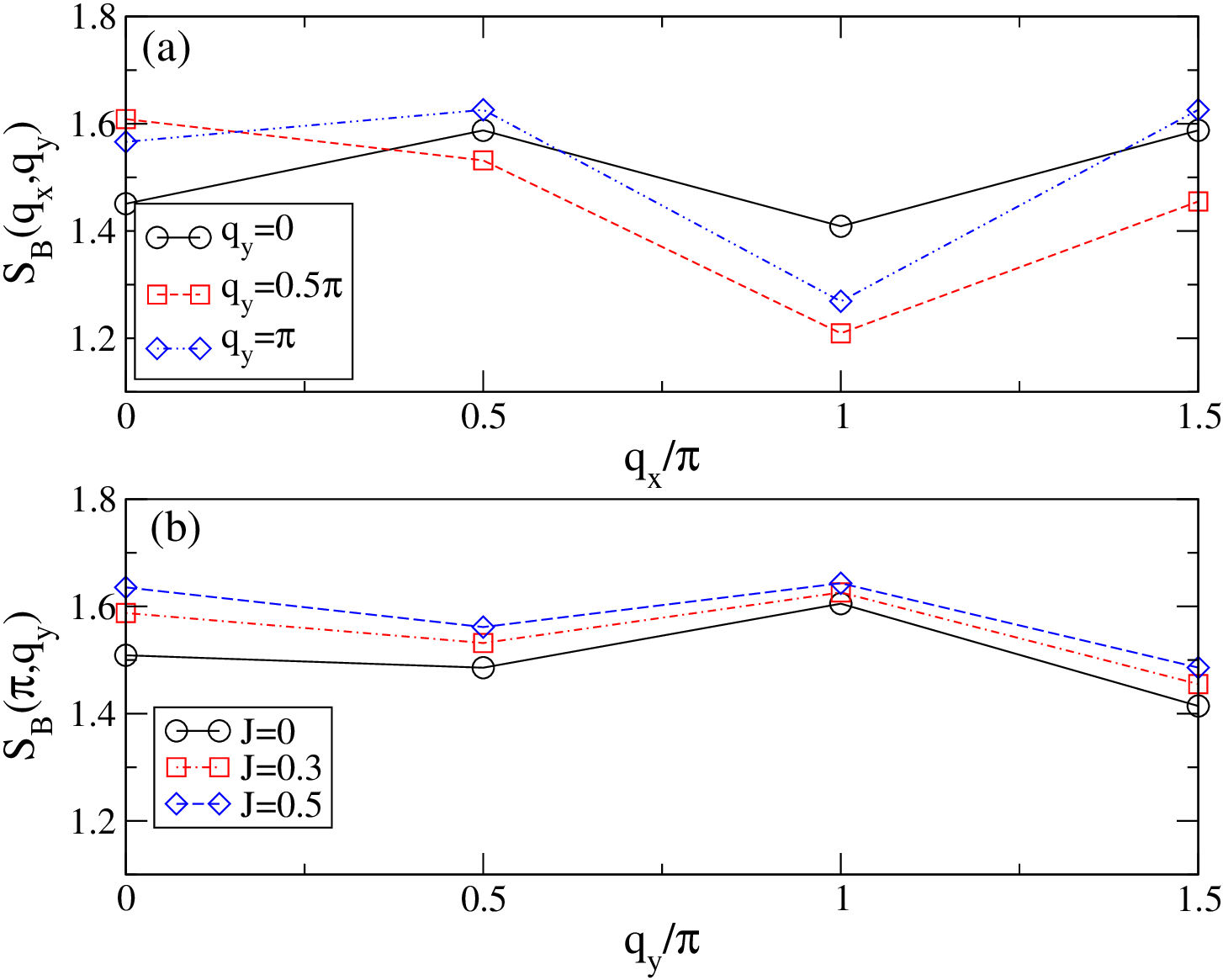}
\caption{(color online) (a) Bond-Bond structure factor
$S_{B_x}(q_x,q_y)$ as a function of $q_x$, $q_y$ for $t'=0.9$, $U=8$, and $J'=J=0.3$. 
(b) $S_{B_x}(\pi,q_y)$ for $U=8$ and varying  $J$. Lines are
only guides to the eye.}
\label{sbofig}
\end{figure}
The VBS order parameter is the bond-bond structure factor 
$S^x_B({\bf Q})$ for bonds along the $x$ axis, defined as
 \begin{equation}
 S^x_B({\bf Q})=\frac{1}{N}\sum_{i,j}e^{i{\bf Q}\cdot{\bf r_{ij}}} \langle 
 (B_{i,i+\hat{x}}-\langle B\rangle) (B_{j,j+\hat{x}}-\langle B\rangle)\rangle,
 \label{bosfac}
\end{equation}
where $\langle B\rangle$ is the expectation value of the bond order.
We consider just the $J'=J$ case, as no NMI phase exists when $J'=0$
(see Fig.~\ref{bofig2}).  VBS order with columnar dimer pattern as
claimed in \cite{Rau11a} would correspond to a peak in $S^x_B({\bf
  Q})$ at ${\bf Q}=(\pi,0)$.  Fig.~\ref{sbofig}(a) shows $S^x_B({\bf
  Q})$ for $U=8$, $t'=0.9$, and $J'=J=0.3$. Throughout the NMI region
($6 \alt U \alt 12$) we find no changes in the ${\bf Q}$-dependence of
$S^x_B({\bf Q})$. As shown in Fig.~\ref{sbofig}(a), all orderings with
$q_x=\pi$, corresponding to bond alternation along $x$, are {\it
  suppressed} compared to other values of $q_x$.  Fig.~\ref{sbofig}(b)
shows the $J$ dependence of $S^x_B(q_x=\pi,q_y)$. As $J$ increases, no
peaks develop; rather the effect of $J$ is simply a renormalization
affecting all $S^x_B({\bf Q})$ at all ${\bf Q}$ equally. Since the VBS
order is absent within the NMI phase for $J=0$ to begin with
\cite{Watanabe03b}, this is strong evidence that the $J$ interaction
does not lead to a VBS phase.

\begin{figure}
\includegraphics[width=3.0in]{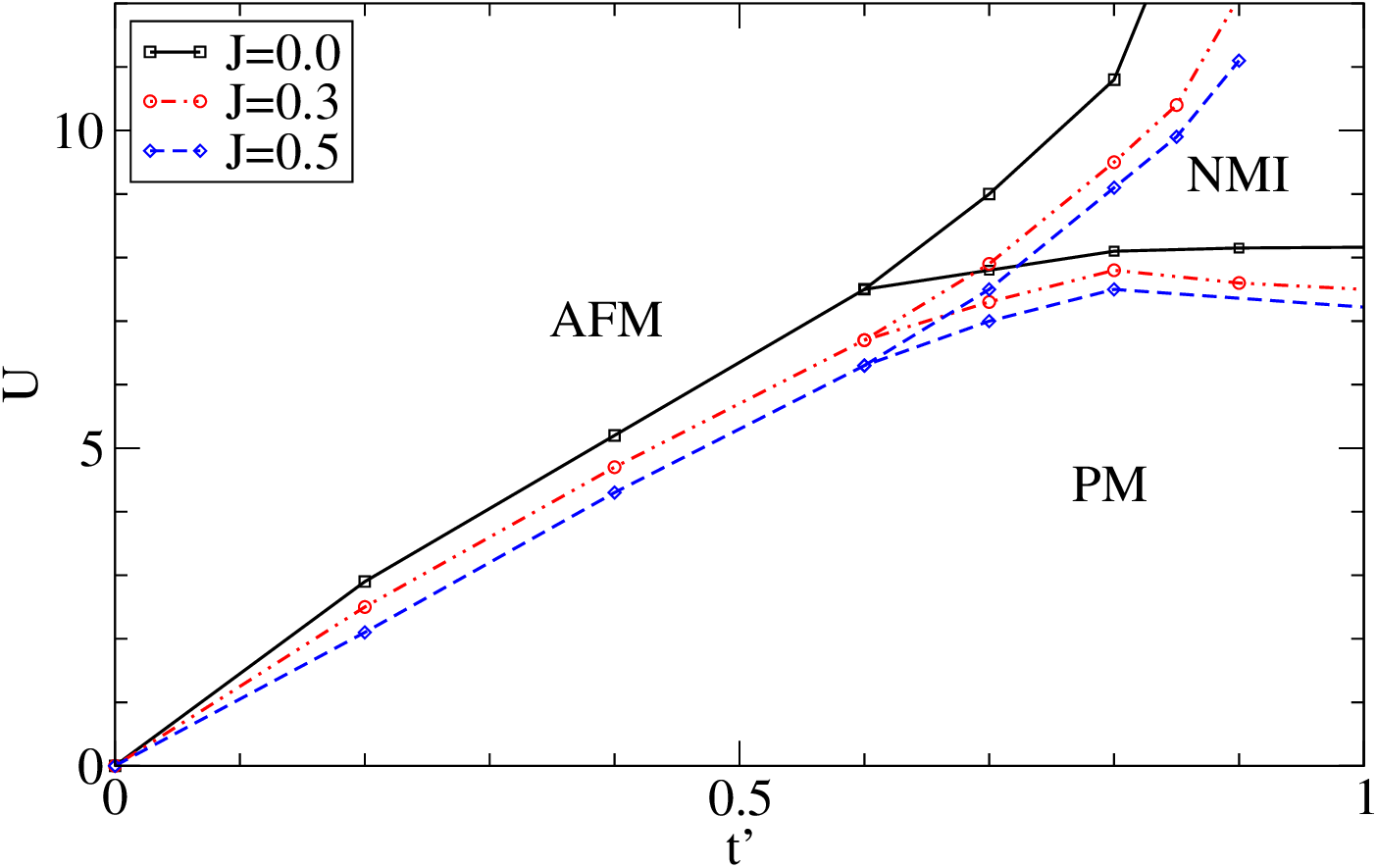}
\caption{(color online) The phase diagram with $J'=J$ for
parameter regions relevant for the dimerized CTS.}
\label{phasediag}
\end{figure}
In summary, the inclusion of Heisenberg exchange interactions in the
effective $\frac{1}{2}$-filled band for for $\kappa$-(ET)$_2X$ and
Z[Pd(dmit)$_2$]$_2$ merely strengthen the AFM phase, and either
eliminates the NMI phase (for $J'=0$) or reduces its width in the
phase diagram (for $J'\neq 0$).  Importantly, neither SC nor VBS
phases are present in the $\frac{1}{2}$-filled band model for any
realistic $J$.  The phase diagram (Fig.~\ref{phasediag}) remains
nearly identical with the phase diagram of the bare
$\frac{1}{2}$-filled Hubbard model \cite{Clay08a,Dayal12a}.  The
absence of SC and VBS suggest that while the approximate methods used
previously can yield classical ordering such as AFM, spin-singlet
formation, a quantum mechanical effect, may be outside the capability
of these techniques.  The question that we started with - what is the
minimal model that describes the CTS - then continues to be
relevant. The only choice appears to be the interacting frustrated
$\frac{1}{4}$-filled band model \cite{Mazumdar11a}, which has the
added advantage that it applies equally well to both the dimerized
$\kappa$-(ET)$_2$X and Z[Pd(dmit)$_2$]$_2$ and the undimerized
$\theta$-(ET)$_2$X which show CO-SC (as opposed to AFM-SC) transitions
(recall that the charge carrier density per molecule is the same in
these two classes of CTS).  In recent work \cite{Li10a,Dayal11a} we
have shown that a frustration-driven AFM-spin singlet transition
occurs within the dimerized interacting $\frac{1}{4}$-filled band
model, where the spin singlet state also exhibits CO. Structural
analysis shows that the so-called VBS state in
EtMe$_3$[Pd(dmit)$_2$]$_2$ has the same CO pattern
\cite{Tamura09a,Yamamoto11a} as that in the coexisting CO-spin singlet
found in the frustrated $\frac{1}{4}$-filled band
\cite{Li10a,Dayal11a}.  Experiments also find charge fluctuations in
the QSL state of $\kappa$-(ET)$_2$Cu$_2$(CN)$_3$
\cite{Abdel-Jawad10a}, suggesting again the applicability of the
$\frac{1}{4}$-filled band model.  We are investigating whether
frustration-induced SC occurs within the latter model.

This work was supported by the Department of Energy grant
DE-FG02-06ER46315.

\end{document}